# GVPT – A software for guided visual pitch tracking

Hyunjin Cho[1,2†], Farhad Tabasi[2†], Jeremy D. Greenlee[2,3*] and Rahul Singh[1*]

Department of Computer Science, University of Iowa, Iowa City, IA[1]

Department of Neurosurgery, University of Iowa Health Care, Iowa City, IA [2]

Iowa Neuroscience Institute, Iowa City, IA [3]

[†] Contributed equally to this work

*Corresponding authors

**Abstract**

GVPT (Guided visual pitch tracking) is a Windows desktop application designed to support speech and voice control experiments, focusing on precise real-time pitch detection. The software offers five distinct pitch patterns (targets), each designed to evaluate different aspects of pitch control, such as pitch sustainability and continuous modulation during speech production. Real-time pitch extraction with minimal latency provides real-time visual feedback to the participant. Additionally, this real-time feedback can be delivered through three visual guidance styles regarding the pitch target: no guidance, guidance at every point, or guidance at key pitch-changing points (contours). After each vocal production trial, an audiovisual playback of the preceding trial can be provided as feedback to the participant.

Voice data recording sampling rates can be customized, producing datasets for thorough post-experiment acoustic analyses. All experimental variables—grading thresholds, guidance types, countdown timers, and pitch change (in cents)—are real-time adjustable. The software also allows the proctor to add or remove tasks from the queue mid-experiment, offering dynamic control. Moreover, GVPT supports transistor-transistor logic (TTL) logging for marking event times at multiple time points. This feature makes this software compatible with behavioral and neuroscience research, including functional neuroimaging and electrophysiological studies, to align the recorded physiological signals with the voice tasks. This software is developed in C# using WPF (Windows Presentation Foundation) and is freely available for download as a Windows executable file.

**Introduction**

Volitional voice control is a fundamental aspect of human spoken communication. This complex behavior requires the coordination of sensory inputs and motor commands to meet communicative goals and produce purposeful actions (Behroozmand et al., 2016; Behroozmand et al., 2015; Zhang, 2016). Sensory information, primarily in the form of feedback, plays a key role in regulating motor commands to adjust vocal output as needed. The integrity of this feedback processing is critical for meaningful and effective communication; disruptions of this sensorimotor integration during vocalization contribute to various clinical conditions, including stuttering, dysarthria, and aphasia, which can significantly affect the quality of life (Hartwigsen & Saur, 2019; Hickok et al., 2011; Sares et al., 2018).

The ability to control voice fundamental frequency ($f_0$; perceived as pitch) is a distinctive feature of human vocal control. Pitch control enables the speaker to convey complex verbal and non-verbal messages. These messages are often accompanied by emotional elements, modulation of speech rhythm and melody (i.e., prosody), or signing. Pitch control is also essential for intelligible and natural speech. It is compromised in a wide range of speech disorders, including apraxia of speech (Odell & Shriberg, 2001), stuttering (Weber-Fox et al., 2013), dysarthria of different etiologies (Patel, 2002), as well as in certain developmental disorders (Peppé et al., 2011). These deficits are thought to involve impairments in sensorimotor processing, affecting the ability to regulate vocal control. Improved understanding of the mechanisms of voice control and learning new vocal patterns is key to advancing knowledge of such disorders and developing effective treatment strategies.

A large body of studies in speech and auditory neuroscience has focused on pitch perturbation. However, the problem of learning new pitch sequences remains largely unexplored. To address this issue, we developed a real-time pitch tracking tool with a visual component (Guided Visual Pitch Tracking; GVPT) that provides real-time visual feedback to monitor performance. GVPT software extracts voice pitch in real-time with minimal delay and generates pitch patterns (targets). This allows the participant to monitor their vocalization relative to the target, and the audiovisual playback of this trial can be played back as a listening trial. GVPT enables us to study the learning processes involved in vocal pitch control using a multimodal feedback approach with high flexibility and precision across different parameters.

**GVPT Software**

    a. *Design philosophy and rationale*

        Multisensory training can optimize and enforce perceptual learning by engaging the full potential of perceptual machinery (Shams & Seitz, 2008). Previous studies have shown that visual clues, as an additional perceptual modality, can facilitate the learning of phonetic features (Xi et al., 2020) as well as intonation patterns and pitch movement (Hazan et al., 2005; Hori et al., 2025) in second language learners.

        Visual pitch tracking was developed as a multisensory feedback paradigm to enforce the learning of novel vocal patterns adapted from tonal language elements (Wang et al., 2022). In this paradigm, the target vocal pattern is visually demonstrated, while the learner's real-time pitch production is superimposed on it, facilitating immediate performance monitoring. This integration of audiovisual information enables a more refined vocal performance and improved pitch modulation.

    b. *Architecture and implementation*

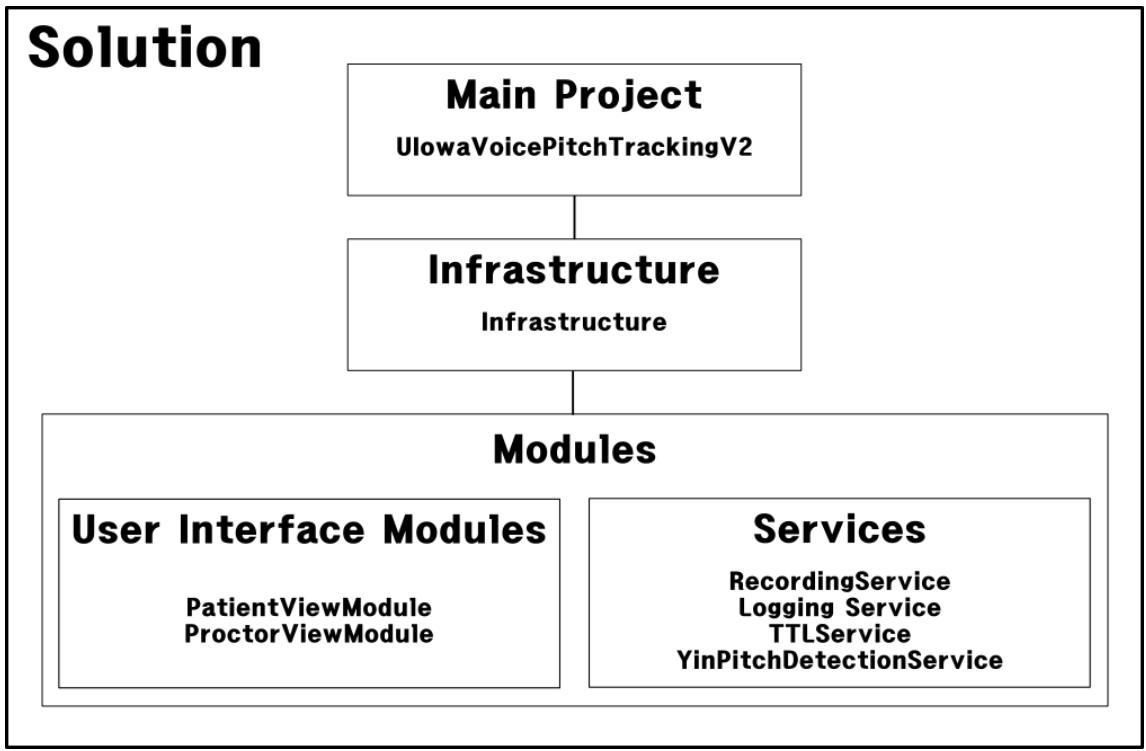

**Figure 1.** *The GVPT software employs the MVVM (Model-View-View-Model) design pattern to achieve high modularity and separation of concerns. In this design pattern, key features are developed as separate modules and interfaced with the main project through the infrastructure layer, ensuring a compact, modular, and scalable software architecture.*

GVPT is developed in C# using Windows Presentation Foundation (WPF) for Windows operating systems. It is designed to support visual pitch tracking paradigms, offering a variety of pitch targets such as sustainment, elevation, lowering, and continuous modulation. By assessing and enhancing a user's ability to visualize their produced vocal pitch, GVPT is a valuable tool for research and training in pitch control.

GVPT also facilitates seamless data analysis through a comprehensive logging system. Each recording session is stored in a dedicated folder, and upon completion, the GVPT software saves the entire audio file from the session. Alongside the audio, it records timestamped logs of detected voice pitches and event logs that detail which tasks were prompted to the user and their completion times. It also stores the base, upper, and lower target pitches for all tasks, as well as the voice pitch contours associated with each task. This structured data collection enhances the feasibility of data analysis.

c. The Proctor Console Interface

The Proctor Console Window of the GVPT software is organized into five essential components, each designed to enhance control and efficiency during auditory pitch training sessions: the Transistor-Transistor Logic (TTL) Console, Recording Control Panel, Recording Parameters, Experiment Parameters, and Visual Pitch Task Controls.

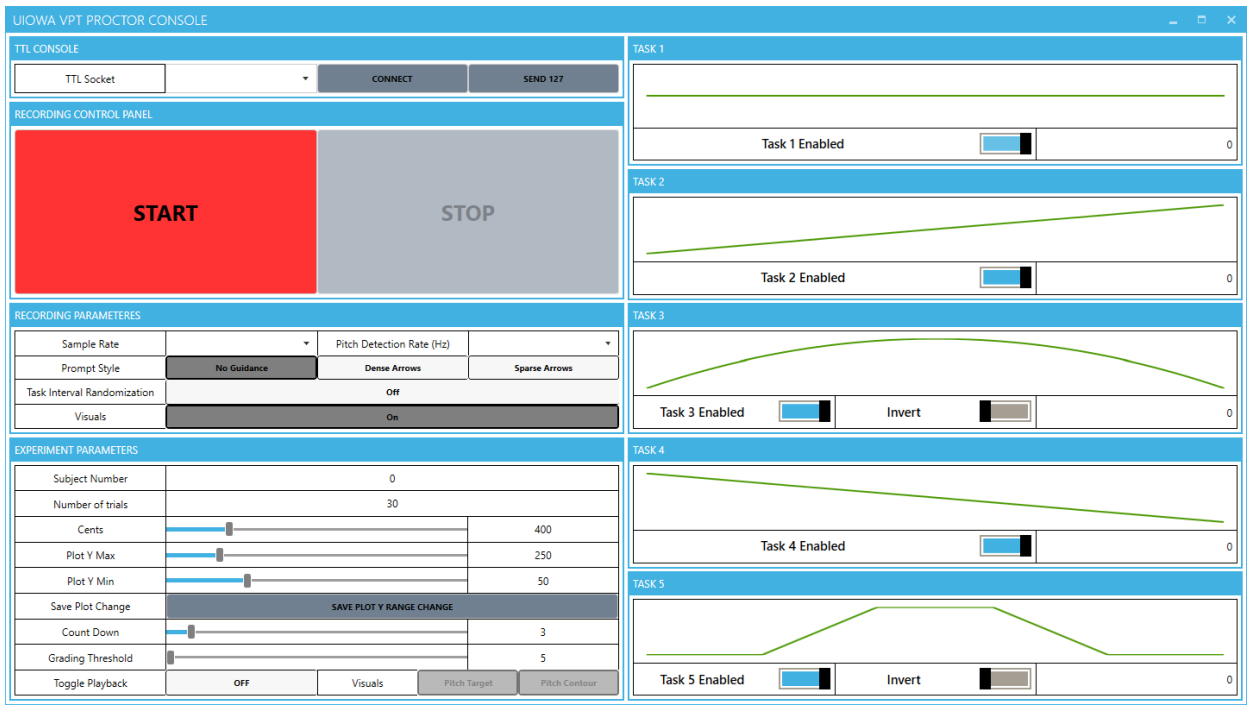

**Figure 2.** *The proctor console window features an intuitive Graphical User Interface (GUI) for the experiment setup.*

- **TTL Console:** The TTL Console enables the proctor to connect a TTL device via an appropriate COM port. This feature is crucial when using a secondary recording machine that needs to synchronize timestamps with the experiment logs, allowing for accurate data alignment and analysis.
- **Recording Control Panel:** This panel facilitates the starting and stopping of experiment sessions, offering the proctor complete control over the session timings. It serves as the main interface for managing the temporal aspects of the training sessions.
- **Recording Parameters:** This component allows users to customize recording settings to suit specific experimental needs. Users can adjust the sample rate, voice pitch detection rate, and user guidance type. It also features options to randomize delay times between tasks and to toggle visual displays, which show detected pitches and pitch targets to subjects, enhancing the interactive training experience.
- **Experiment Parameters:** The proctor can input critical experiment-specific details, such as the subject number, which helps in organizing logs, selecting the number of trials, and setting the voice pitch-shifting amount in cents. It includes controls for adjusting the Y-axis to match the subject's base pitch, setting the countdown timer for trials, and changing the grading threshold. Additionally, a toggle for post-trial playback allows the user to review the performed pitch points from the previous trial relative to the pitch target, facilitating passive audiovisual performance feedback.
- **Visual Pitch Task Controls:** located on the right side of the Proctor Console Window, controls visual pitch tasks (targets) with toggle switches that allow the proctor to activate or deactivate specific voice pitch targets, providing flexibility in customizing the training sessions. Additionally, the switches enable the inversion of pitch targets - changing an 'up' task to a 'down' task. A counter is displayed within this section, which tracks the number of trials each task has been performed in the current experiment. The current version does not support custom pitch contours designed by users.

### d. The Subject Interface

The Subject Window serves multiple functions essential for facilitating pitch production and control. It displays a countdown timer and a 'go' cue to initiate visual pitch tracking tasks and real-time voice pitch detection. During the visual pitch tracking task, subjects are instructed to maintain their voice at their usual/normal base pitch. Following a 1.5-second interval after the 'go' cue, GVPT assesses the subject's current voice pitch and calculates and displays the visual pitch target. Performance qualitative feedback is visually delivered through emoticons—a smiley for successful alignment with pitch targets when the subject tracked the target between a grading threshold of more than 75% of the task duration, a neutral face for close attempts (between 25% and 75%), and an angry face for significant deviations (less than 25% of the task duration). Additionally, if the proctor enabled post-trial playback to allow subjects to listen and watch their performance passively after each try, the voice pitch target and the subject pitch contour are reconstructed (audiovisual playback).

The proctor can choose from various guidance prompts for the subject to highlight discrepancies. Options include no additional visual guidance; dense arrow guidance at each pitch point (every 25ms or 50ms), where red arrows are drawn from each detected pitch point outside the grading boundary toward the target; or sparse arrow guidance, which provides arrows only during moments of pitch change (contours) in the voice pitch task.

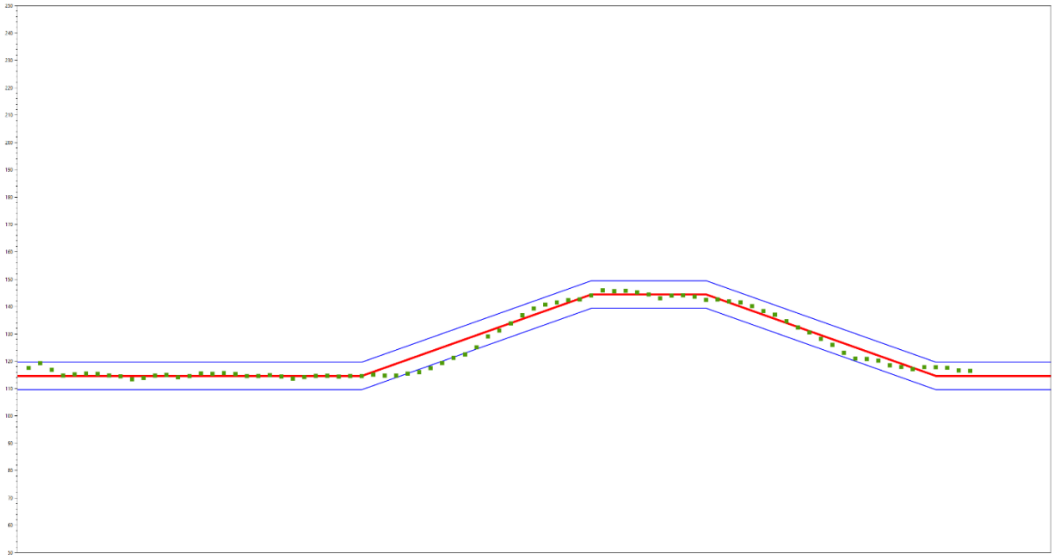

**Figure 3**. *The red line represents the pitch target, while the blue lines indicate the grading boundaries. Detected subject voice pitches are marked as green squares. This visual setup helps the subject adjust their pitch in real time to align more closely with the target, providing immediate feedback on performance.*

e. Core features

The GVPT application is designed to enhance visual pitch tracking through several key functionalities:

1. **Real-Time Pitch Tracking and Display:** GVPT continuously tracks the subject's voice pitch in real time using the YIN fundamental frequency estimator (de Cheveigné & Kawahara, 2002) and displays this information dynamically—the proctor has the option to detect and display the subject's pitch every 25ms or 50ms. This immediate feedback allows subjects to adjust their pitch in line with the visual targets provided, thereby facilitating more effective training and promoting faster improvement.

2. **Real-Time Experiment Modification:** The proctor can adjust the difficulty of the visual pitch tasks (e.g., pitch change amount, changing the task to an easier task) based on the subject's performance during the recording session. This adaptive approach ensures that the training always aligns with the subject's current capabilities, pushing for improvement while maintaining engagement.

3. **Comprehensive Logging System:** GVPT features a thorough logging system that organizes data from each session into individual folders named according to the subject ID provided in the Experiment Parameters interface. This system captures detailed information, including the complete audio file of the session, timestamped voice pitch logs, and event logs that document the sequence and completion of tasks, as well as every task base and target pitch and the pitch contour of the target. Such detailed data collection is crucial for subsequent analysis, allowing researchers or trainers to assess progress over time and make informed decisions about future training approaches.

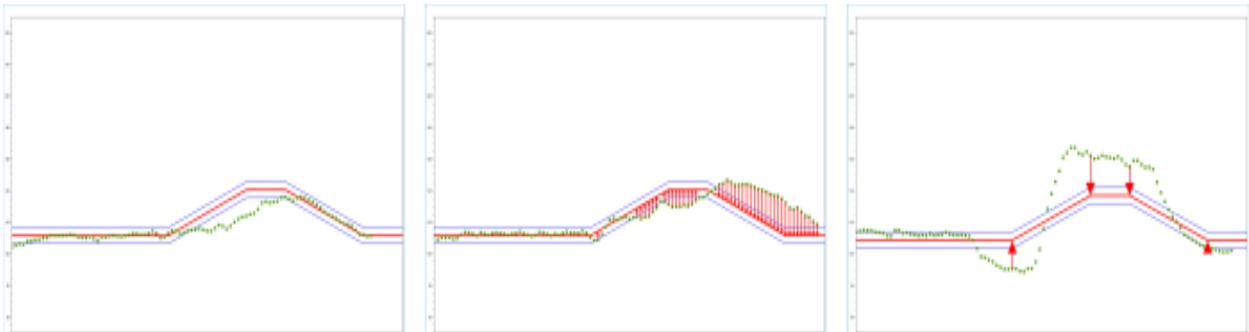

**Figure 4**. *Three guidance prompts are available in the Subject Window. The left panel shows no additional guidance; the center panel displays dense guidance, where multiple red arrows indicate discrepancies from the target pitch that are outside of the grading boundary; the right panel features sparse guidance, providing arrows only at critical pitch change moments, simplifying the visual feedback.*

*f. Software System Architecture and Implementation.*

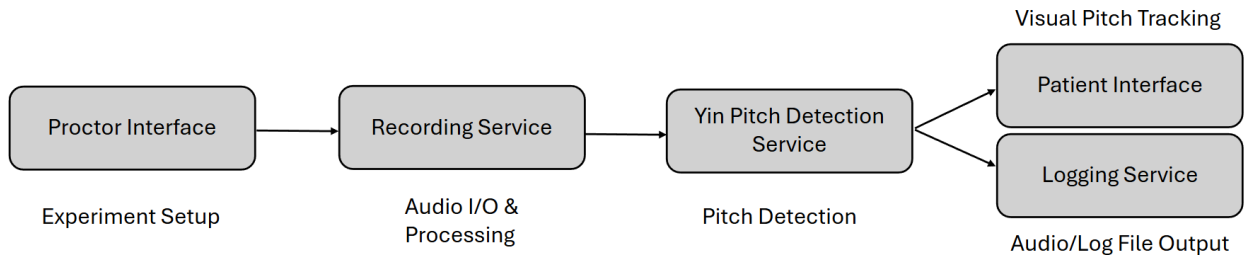

**Figure 5**. *The operation flow of GVPT: The experiment is set up in the Proctor Interface, and the Recording Service is initiated to manage audio I/O and processing. Once audio is captured, the Yin Pitch Detection Service extracts the pitch frequency from the subject's voice. The detected pitch information is then delivered to the Patient Interface for visual pitch tracking and to the Logging Service, which records audio, detected pitch, and task logs for future analysis.*

The GVPT interface utilizes MahApps.Metro v1.6.5 (.NET Foundation & Contributors, 2018) to provide a modern look. For advanced and real-time plotting in the Subject Window, OxyPlot v2.1.2 (Bjørke et al., 2022) is employed. Audio functionalities such as recording, pitch detection, and playback are managed using NAudio v2.2.1 (Heath et al., 2023). Task management is streamlined with MoreLINQ v4.3.0 (Aziz et al., 2024), while NLog v5.2.5 (Kowalski et al., 2023) supports a robust logging module. Finally, Prism.Mef v6.3.0 (Lagunas & Siegel, 2017) and ReactiveProperty v9.5.0 (Kawai et al., 2024) are integrated to adhere to the MVVM (Model-View-ViewModel) design pattern, a cornerstone of Microsoft's C# programming practices, ensuring a compact and maintainable code base.

*g. Technical User Guidebook*

A comprehensive technical user guidebook is available to assist users in operating the GVPT software. It can be accessed at

https://dupont.cs.uiowa.edu/software/brain/gvpt/files/UIowa_GVPT_User_Manual.pdf

**Software and Code Availability:**

The executable file and its C# code base can be downloaded at

https://dupont.cs.uiowa.edu/software/brain/gvpt/gvpt.html

**Conclusion**

The GVPT tool is designed to detect and visually represent voice pitch in real time, allowing users to monitor their pitch modulation relative to a target with precision. GVPT is particularly valuable for evaluating pitch production, learning new vocal patterns, and adjusting dynamic pitch. When paired with simultaneous brain recordings, such as electrophysiology or functional magnetic resonance imaging (fMRI), GVPT provides a versatile and powerful means to investigate the neural mechanisms of how the brain coordinates sensorimotor information for vocal behaviors.

**Acknowledgements**:

We thank Mathew A. Howard, Christopher I. Petkov, Matthew R. Hoffman, Roozbeh Behroozmand, and Haiming Chen for their support, as well as all pilot subjects who provided constructive feedback on the software.